%% file: WSC2020Template.tex
\documentclass{wscpaperproc}
\usepackage{latexsym}
\usepackage{graphicx}
\usepackage{mathptmx}
\usepackage{enumitem}
\usepackage[mathscr]{euscript}
\usepackage[ruled,noline]{algorithm2e}
\usepackage{subcaption}
\usepackage{amsmath}
\usepackage{amsfonts}
\usepackage{amssymb}
\usepackage{amsbsy}
\usepackage{amsthm}
\usepackage{bm}
\usepackage{xspace}
\usepackage{framed}
\usepackage[pdftex,colorlinks=true,urlcolor=blue,citecolor=black,anchorcolor=black,linkcolor=black]{hyperref}
\newtheorem{assumption}{Assumption}

\newtheorem{example}{Example}

\hypersetup{colorlinks = true, urlcolor = blue, linkcolor = blue,
  citecolor = blue }
\newcommand{\xmath}[1]{\ensuremath{#1}\xspace}

\newcommand{\RV}{\xmath{\mathscr{RV}}}

\newcommand{\R}{\xmath{\mathbb{R}}}
\newcommand{\mv}[1]{\bm{#1}}

\newcommand{\qq}{\mv{q}}

\newcommand{\xx}{\mv{x}}
\newcommand{\XX}{\mv{X}}
\newcommand{\YY}{\mv{Y}}
\newcommand{\ZZ}{\mv{Z}}
\newcommand{\yy}{\mv{y}}

\newcommand{\pp}{\mv{p}}
\newcommand{\e}{\mathrm{e}}
\newcommand{\Lev}{\xmath{\textnormal{lev}}}

\newcommand{\Real}{\mathbb{R}}
\newcommand{\Prob}{P}
\newcommand{\Expc}{\mathbb{E}}

\newcommand{\fLD}{f_{\textnormal{LD}}}
\newcommand{\supp}{\textnormal{supp}}
\newcommand{\cis}{\hat{C}_{\beta,n}^{\textnormal{IS}}}
\newcommand{\vis}{\hat{v}_{\beta,n}^{\textnormal{IS}}}

\newcommand{\CVaR}{\text{CVaR}}
\newcommand{\VaR}{\text{VaR}}

\newtheoremstyle{wsc}
{3pt}
{3pt}
{}
{}
{\bf}
{}
{.5em}
{}

\theoremstyle{wsc}
\newtheorem{theorem}{Theorem}

\begin{document}

%
%

\pagestyle{fancyplain}

\thispagestyle{plain}
\firstPageHead{}

\chead{\fancyplain{}{\itshape Deo, Murthy}}

\rhead{}
\cfoot{}
\renewcommand{\headrulewidth}{0pt} 

\input{wscbib.tex}           

\setlength{\baselineskip}{12.7pt}

\title{EFFICIENT BLACK-BOX IMPORTANCE SAMPLING FOR \VaR\ AND \CVaR \ ESTIMATION}

\author{Anand Deo\\[12pt]
Singapore University of Technology and Design\\
8 Somapah Rd, Singapore 487372
\and
Karthyek Murthy\\[12pt]
Singapore University of Technology and Design\\
8 Somapah Rd, Singapore 487372
}

\maketitle

\section*{ABSTRACT}
This paper considers Importance Sampling (IS) for the estimation of tail risks of a loss defined in terms of a sophisticated object such as a machine learning feature map or a mixed integer linear optimization formulation. Assuming only black-box access to the loss and the distribution of the underlying random vector, the paper presents an  efficient IS algorithm for estimating the Value at Risk and Conditional Value at Risk. The key challenge in any IS procedure, namely, identifying an appropriate change-of-measure, is automated with a self-structuring IS transformation that learns and replicates the concentration properties of the conditional excess from less rare samples. The resulting estimators enjoy asymptotically optimal variance reduction when viewed in the logarithmic scale. Simulation experiments highlight the efficacy and practicality of the proposed scheme.

\section{INTRODUCTION}
\label{sec:intro}
Value at Risk (\VaR) and Conditional Value at Risk (\CVaR) constitute two widely used measures of tail risk in quantitative risk management \cite{mcneil2015quantitative}. Desirable properties such as subaddivity, convexity, etc., have allowed \CVaR \ to   further flourish as a vehicle for introducing risk aversion in planning problems in operations and machine learning (see, for example, \shortciteNP{rockafellar2000optimization,bienstock2014chance,ban2018machine,tamar2016sequential}). The value at risk at a tail probability level $\beta$ is the $(1-\beta)$-th quantile of the loss distribution. \CVaR \ at level $\beta$ is the expected value of the loss over its largest $\beta$ fraction of outcomes and is relatively more challenging to estimate than \VaR \ \cite{lim2011conditional}. 
With a limited fraction of data representing the loss distribution tail, estimation of \VaR/\CVaR\ via simulation is executed typically with a rare event simulation technique such as importance sampling, splitting, conditional Monte Carlo or control variates for the purposes of variance reduction and accelerated estimation \cite{glasserman2013monte}. 

As we explain imminently in the context of importance sampling (IS), efficient use of these simulation techniques rely often on leveraging the structure of the loss at hand and the distribution of the underlying random vector. In terms of general methodology, \citeNP{glynn1996importance} demonstrates how variance reduction using IS in tail estimation can be translated to efficient estimation of \VaR. \citeNP{sun2010asymptotic} develop asymptotic representations for \VaR  /\CVaR \ which yield conveniently applicable characterizations of asymptotic variances for \VaR \ and \CVaR. \shortciteNP{bardou2008computation,egloff2010quantile} and, more recently, \shortciteNP{he2021adaptive} develop adaptive algorithms which incorporate generic IS changes of measure in estimation of \VaR/\CVaR. While generically applicable, it is not within the scope of these works to provide specific prescriptions of IS changes of measure that offer variance reduction guarantees. In this regard, \shortciteNP{glasserman2000variance,glasserman2002portfolio,BJZWSC} demonstrate how the properties of multivariate normal and $t$ distributions can be exploited to reap substantial variance reduction in portfolio risk estimation contexts. These algorithms critically utilize the specific structural properties of the loss, such as the linear-quadratic or the sum of indicators structure, and are restricted to settings involving multivariate normal and $t$ distributions.

 In a number of operations and risk management contexts, 
 the underlying loss often involves a sophisticated structure. Planning problems typically specify a loss  in terms of an optimization formulation involving numerous constraints. In the rapidly growing instances of operations and risk management models which use machine learning tools, a suitable loss is written in terms of a feature-map (or) a feature-based decision rule specified, for example, in terms of representation-learning devices such as kernels or deep neural networks (see \shortciteNP{ban2019big,SmartPredictOpt} and references therein). Given the rich modelling power of these loss instances, it is impractical to explicitly tailor the IS change of measure to the problem considered.  Adaptive IS methods, which utilize the estimator variance (or) cross-entropy criterion \cite{rubinstein2013cross} to search for the best parameter choice within a chosen IS distribution family, remain the most common approach to address this challenge. The performance of the adaptive approaches is  however determined crucially by the IS family distribution family initially chosen and may additionally involve systematic underestimation \shortcite{arief2021deep}. 
 
 This paper aims to tackle the challenges in marrying efficiency with black-box IS for \VaR/\CVaR\ estimation.  Restricting to multivariate normal distributions, \shortciteNP{bai2020rare,arief2021deep} utilise the machinery of dominating points to algorithmically arrive at efficient IS mixture distributions for estimation of distribution tails of losses that can be either  directly written or approximated with a piece-wise linear structure.
 Assuming only a black box access to the evaluations of loss $L(\cdot)$ and the distribution of the underlying random vector $\XX,$  we present here an efficient IS algorithm (Algorithm~\ref{algo:CVaR-I.S.}) to jointly estimate \VaR /\CVaR \  of $L(\XX)$. 
The IS scheme in this paper builds upon a generically applicable large deviations framework and the IS scheme developed in \citeNP{deo2021achieving} for the estimation of distribution tails. Exploiting the self-similarity in conditional excess distributions at different thresholds, the novel approach informs a suitable IS measure by extrapolating excess loss samples observed at less rare thresholds. We show that the proposed IS scheme offers asymptotically optimal variance reduction, when viewed at a logarithmic scale,  for a broad class of useful losses and multivariate distributions.  Specifically, given any $\varepsilon > 0,$ we show that the sample complexity for estimating \CVaR \  at a tail probability level $\beta$ scales as $O(\beta^{-\varepsilon})$ with the proposed IS scheme. It is instructive to contrast this with the scaling of $O(\beta^{-1-\varepsilon})$ obtained for the case of naive estimation without IS.  We complement the variance reduction guarantees with numerical experiments that validate the efficacy and generic applicability of the proposed scheme. 
 
 We note that an attempt at black box \CVaR\ estimation is made by \citeNP{deo2020optimizing} for the case where $\XX$ has regularly varying tails (that is, when $P(X_i>x)\sim x^{-\alpha_i}$, for $\alpha_i>0$). While their scheme bears some similarity to Algorithm~\ref{algo:CVaR-I.S.}, it relies heavily on the weak convergence properties of regularly varying densities, and does not result in asymptotically
    optimal variance reduction.

\noindent \textbf{Notation:} 
We use $\xrightarrow{\mathcal{D}}$ to denote convergence in distribution. Boldface letters denote vectors. 
Likewise for a function $\mv{f}:\R^d\to\R^k$, $\mv{f}(\xx) = (f_1(\xx),\ldots, f_k(\xx))$.  We let $N(\mu,\sigma^2)$ denote a normal variable with mean $\mu$ and variance $\sigma^2$.  Let $\|\xx\|_p $ denote the $\ell_p$ norm of a vector $\xx\in \R^d$ and  
 $B_r(\mv{x})$ denote the $l_\infty$-metric ball of radius $r$ centred at $\xx$. For an increasing function $f:\Real \rightarrow \Real$, we let $f^{-1}$ denote its left-inverse.  For real valued functions $f$ and  $g$, we say that $f(x)=O(g(x))$ as $x\to\infty$ if there exist positive constants $M,x_0$ such that for all $x>x_0$, $|f(x)|\leq M|g(x)|$. We say that $f(x)=\tilde{O}(g(x))$ if $f(x)=O(g(x) \log^k(x)),$ for some $k>0$.  
\section{PROBLEM DESCRIPTION}\label{sec:Setup}
Suppose $L(\xx)$ denotes the loss incurred when the underlying random vector $\XX$ realizes the value $\xx$. 
Let $F_{L}$ denote the distribution function of $L(\XX)$, that is, $F_L(u) = \Prob(L(\XX) \leq u)$, and let $f_L$ be its density. Given a confidence level $\beta\in (0,1)$, denote the Value at Risk (\VaR) and Conditional Value at Risk (\CVaR) of the loss $L$ at level $\beta$ as, 
\[
v_{\beta}  = F_L^{-1}(u) := \inf\{u \in \R: F_L(u) \geq 1-\beta\}, \quad \text{and} \quad C_{\beta} := v_{\beta} + {\beta^{-1}}\Expc\left(L(\mv{X}) - v_{\beta}\right)^{+}
\] 
respectively. Our objective is to enable efficient estimation of the \VaR\  $v_\beta$ and \CVaR\  $C_\beta$ for values of $\beta$ close to 0. Assumption~\ref{assume:V} below imposes a mild regularity condition on the function $L(\cdot),$  whose evaluation may be available only via a black-box. 
\begin{assumption}
  \textnormal{The function $L : \Real^d \rightarrow \Real$ satisfies
    the following conditions:
    \begin{enumerate}[label=(\roman*)]
    \item \label{item:1a)} the set $\{\xx \in \supp(\XX): L(\xx) > u\}$ is
      contained in $\Real^d_+$ for all sufficiently large $u;$ and 
    \item \label{item:1b)}for any sequence $\{\xx_n\}_{n \geq 1}$ of
      $\mathbb{R}^d_+$ satisfying $\xx_n \rightarrow \xx,$ we have 
  \begin{align*}
    \lim_{n \rightarrow \infty} \frac{L(n\xx_n)}{n^\rho}= L^\ast(\xx),
  \end{align*}
  where $\rho$ is a positive constant and the limiting function
  $L^\ast: \mathbb{R}^d_+ \rightarrow \mathbb{R}$ is such that the
  cone $\{\xx \in \R^d_+: L^\ast(\xx) > 0\}$ is nonempty.
\end{enumerate}
}
  \label{assume:V}
\end{assumption}
Assumption \ref{assume:V} stipulates that the loss incurred, denoted by $L(\XX),$ is large when at least one of components of $\XX$ takes large values.  Besides commonly considered examples such as  piecewise affine and linear-quadratic losses, Assumption \ref{assume:V} is satisfied for a wide-class of operations and quantitative risk management models that motivate our study. These include cases where $L(\cdot)$ is written as the value of a suitable mixed integer linear program or a quadratic program, and instances in prescriptive analytics where a suitable $L(\cdot)$ is written in terms of  feature maps or decision-rules specified by a neural network with ReLU activation units. We refer the reader to \citeNP[Section 2]{deo2021achieving} for a precise description of these examples for which Assumption \ref{assume:V} is readily satisfied. 
Notably, Assumption \ref{assume:V} does not  require the loss to be convex or possess specific combinatorial structure. 

\noindent \textbf{Monte-Carlo estimation without any change of measure.} 
Given $n$ independent samples $\XX_1,\ldots, \XX_n$ of $\XX$, let $\hat{F}_{n,L}$ denote the empirical cumulative distribution function (c.d.f.) formed from the samples $L(\XX_1),\ldots, L(\XX_n)$. 
Then the \VaR \ and \CVaR \ at level $\beta$ can be estimated as,
\begin{equation}\label{eq:SA-CVaR}
\hat{v}_{\beta, n} = \hat{F}_{n,L}^{-1}(1-\beta) \quad \text{and} \quad \hat{C}_{\beta,n} = \hat{v}_{\beta,n} +
  [n\beta]^{-1} \sum_{i=1}^n \left[ L(\mv{X}_i)
  - \hat{v}_{\beta,n}\right]^+,
\end{equation}
respectively. These estimators  satisfy asymptotic normality under nominal assumptions (see, for example, \citeNP[pg. 75]{serfling2009approximation} and \shortciteNP[Theorem 2]{trindade2007financial}):
\begin{equation}\label{eqn:VaR-CLT}
\sqrt{n}(v_\beta-\hat{v}_{\beta,n}) \xrightarrow{\mathcal{D}}  \sigma_{v}(\beta)N(0,1) \quad \text{and}\quad  \sqrt{n}(C_{\beta} -\hat{C}_{\beta,n})\xrightarrow{\mathcal{D}} \sigma_c(\beta) N(0,1)
\end{equation} 
where 
\begin{equation}\label{eqn:SAA}
\sigma_{v}^2(\beta)  = {\beta(1-\beta)}[f_{L}(v_\beta)]^{-2} \quad \text{and} \quad  \sigma^2_c(\beta) = {\beta^{-2}}\mathrm{Var}\left[\left(L(\XX) - v_{\beta}\right)^+\right].
\end{equation}
The asymptotic variances indicate the price paid in terms of sample complexity when $\beta \searrow 0.$ Observe that \eqref{eqn:VaR-CLT} and \eqref{eqn:SAA} imply that with the error in CVaR estimation with $n$ samples is roughly $N(0,n^{-1}\sigma_c^{2}(\beta))$. It can be seen that $\sigma^{2}_c(\beta) = \tilde{O}(\beta^{-1})$ (see for example, \eqref{eqn:HighVar}). Therefore, estimating    $C_\beta$ within a relative error of $\varepsilon$ with $(1-\delta)$ confidence necessarily requires
$\tilde{O}(\beta^{-1}\delta^{-1}\varepsilon^{-2})$ samples of $\boldsymbol{X}$ when using the above sample-average based estimators;
see also \citeNP{sun2010asymptotic}. Since this sample requirement is impractically large when $\beta$ is small, importance sampling is typically considered in order  to reduce mean square error (MSE) to a lower order than $\tilde{O}(\beta^{-1}).$

\section{THE PROPOSED IS ALGORITHM}
\label{sec:IS-Algo}
We begin by describing the IS scheme presented in Algorithm \ref{algo:CVaR-I.S.} below. To circumvent the issue of limited relevant observations in tail exceedance events of the form $\{L(\XX) > u\},$  
IS typically involves obtaining samples from an alternate distribution under which these exceedance events are less rare. To accomplish this in our context, define the $\Real^d$-valued function $\mv{T}(\xx) := \xx [r_\beta]^{\mv{\kappa}(\xx)}$, where $r_\beta :[0,1)\to \R_+$ is a decreasing function of $\beta$ explicitly identified in Algorithm \ref{algo:CVaR-I.S.} and \begin{equation}\label{eqn:kappa-extrp}
\mv{\kappa}(\xx) := \frac{\log (1+|\xx|)}{\rho\|\log(1+|\xx|)\|_{\infty}}.
\end{equation}
Exponentiation is done component-wise in the above expression for $\mv{T}(\xx)$ as in, $\mv{T}(\xx) = (x_1 r_\beta^{\kappa_1(\xx)}, \ldots,x_d r_\beta^{\kappa_d(\xx)}).$ 
In Algorithm \ref{algo:CVaR-I.S.}, we use independent samples of  $\ZZ := \mv{T}(\XX)$ as the samples from IS distribution specified implicitly via $\mv{T}.$ The map $\mv{T}:\Real^d \rightarrow \Real^d$ can be shown to be invertible almost everywhere on $\Real^d$ (see \citeNP[Proposition 1]{deo2021achieving}) and the resulting vector $\ZZ$ has a probability density if $\XX$ has a density. 
Letting $f_{\XX}$ and $f_{\ZZ}$ denote the respective densities of $\XX$  and $\ZZ,$ the likelihood ratio resulting from this change-of measure is given by, 
\begin{align}
\mathcal{L}_{R} = \frac{f_{\XX}(\ZZ)}{f_{\ZZ}(\ZZ)} = \frac{f_{\XX}(\ZZ)}{f_{\XX}(\XX)} J(\XX) 
\label{eq:LR}
\end{align}

An explicit expression of the Jacobian, $J(\xx) = \partial \mv{T}(\xx) /\partial \xx$ in the above expression,  is given in Algorithm \ref{algo:CVaR-I.S.}.  
With this change-of-measure, we have the following unbiased estimator for the c.d.f. $F_L(u):$
\begin{equation}\label{eqn:FIS}
\hat{F}_{n,L}^{\mathrm{IS}}(u)  = 1-\frac{1}{n} \sum_{i=1}^n  \frac{f_{\XX}(\ZZ_i)}{f_{\XX}(\XX_i)} J(\XX_i) \ \mathbf{I}(L({\ZZ_i}) > u), 
\end{equation}
where ${\XX}_1,\ldots,{\XX}_n$ are drawn i.i.d. from $\XX$, and $\ZZ_i=\mv{T}(\XX_i)$.  Subsequent IS estimation of $v_\beta,C_\beta$ involves a routine computation of \VaR \  and \CVaR \ from the given IS estimator $\hat{F}_{n,L}^{\mathrm{IS}}(u)$ for the c.d.f. and is  described precisely in Algorithm~\ref{algo:CVaR-I.S.} below.

\begin{algorithm}[h]
 \caption{Importance Sampling Algorithm for joint computation of \VaR \ and \CVaR}\label{algo:CVaR-I.S.}
  \textbf{Input:} Target tail probability level $\beta$,  hyper-parameter  $h>0$, $n$ i.i.d. samples 
  $\boldsymbol{X}_1,\ldots,\boldsymbol{X}_n$ from $f_{\XX}(\cdot)$.\\
  \textbf{1. Transform the samples:} For each sample
  $i=1,\ldots,N,$ compute the transformation,
  \begin{align*}
    \ZZ_i = \mv{T}(\XX_i) := \XX_i [r_\beta]^{\kappa(\XX_i)},
   \end{align*}
   where $r_\beta=h \log\log (1/\beta)$ and $\mv{\kappa}(\xx)$ is given as in \eqref{eqn:kappa-extrp}.\\ \textbf{2. Compute the associated likelihood:} For each transformed
   sample $\ZZ_i,$ compute the respective likelihood ratio as,
  \begin{align}
    \mathcal{L}_{R,i} := \frac{f_{\XX}(\ZZ_i)}{f_{\XX}(\XX_i)}  J(\XX_i) 
    \qquad i = 1,\ldots,N, 
    \label{LLR}
  \end{align}
  where $f_{\XX}(\cdot)$ is the density of $\XX$ and
  $ J(\cdot) $ is the Jacobian of the transformation
  $\mv{T}(\cdot)$ given by,
   \begin{align}
    \label{eqn:Jac}
    J(\xx)  &:= \left[\prod_{i=1}^d \tilde{J}_i(\xx) \right]\times \frac{r_\beta^{\mv{1}^\intercal \mv{\kappa}(\xx)}}{\max_{i=1,\ldots,d} \tilde{J}_i(\xx)},\\
    \text{where } \tilde{J}_i(\xx)
            &:= 1+\frac{\rho^{-1}\log(r_\beta)}{\Vert\log(1+|\xx|)
              \Vert_\infty}\frac{|x_i|}{1+|x_i|}, \quad i = 1,\ldots,d.
              \nonumber
  \end{align}
\noindent \textbf{3 Compute the IS based {\VaR \ and \CVaR}:}
   \begin{equation}\label{eqn:CVaR-comp-Is}
       \cis  := \vis + \frac{1}{n\beta}
      \sum_{i=1}^n \big(L({\mv{Z}}_i) - \vis \big)^+
      \mathcal{L}_{R,i},
    \end{equation}
    where IS based \textnormal{\VaR},
    $\vis := \inf\{ u : \hat{F}_{n,L}^{\mathrm{IS}} (u) \geq 1-\beta \},$ is
    estimated from the c.d.f. estimate $\hat{F}_{n,L}^{\mathrm{IS}}(\cdot)$ in
    \eqref{eqn:FIS}.
  \end{algorithm}  

A key feature of Algorithm \ref{algo:CVaR-I.S.} is that it is agnostic to  the specific forms of both the loss $L(\cdot)$ and the distribution of $\XX$ and it requires only a black-box access to evaluations of $L(\cdot)$ and $f_{\XX}(\cdot).$   This is in sharp contrast to most existing literature requiring careful tailoring of the IS density to the underlying distribution and the loss considered. 
Building on the self-structuring IS procedure introduced in \citeNP{deo2021achieving} for estimating tail probabilities of the form $P(L(\XX) > u),$ Algorithm~\ref{algo:CVaR-I.S.} below offers a suitable adaptation to the root-finding task required to estimate \VaR. In contrast to estimating $P(L(\XX) > u)$ for a fixed large $u$, VaR/\CVaR\  estimation requires that the extrapolation parameter $r_\beta$ 
is chosen carefully as a function of $\beta$ such that variance reduction is pronounced even if the precise range of $u$ over which root-finding has to be conducted for quantile estimation is not known apriori. The choice of  hyperparameter $h$ can be made either with a cross-validation based approach we demonstrate in numerical experiments, or with  recursive schemes such as those considered in \citeNP{bardou2008computation} or \shortciteNP{he2021adaptive}.


\section{VARIANCE REDUCTION GUARANTEES FOR ALGORITHM~\ref{algo:CVaR-I.S.}}  \label{sec:algo} 
Let $\mv{\Lambda}(\xx)= (\Lambda_1(x_1),\ldots, \Lambda_d(x_d))$, where $\Lambda_i(x) = -\log \Prob(X_i \geq x)$ denotes the hazard function of component $X_i.$  We say that $f:\Real \rightarrow \Real$ is regularly varying if for all $x\in \R_+$, 
\[
\lim_{t \rightarrow \infty} \frac{f(tx)}{f(t)} = x^{p}, 
\]
for some $p \in \Real$ (see \citeNP[Definition B.1.1]{de2007extreme}). In this case, we write $f \in \RV(p).$ 
Letting $\YY := \mv{\Lambda}(\XX),$ we see that vector $\YY$ has standard exponential distribution as marginals. Just as in the use of copula models, standardization of marginals allows to state the main result without getting distracted by the differing marginal distributions.
\begin{assumption}
\label{assume:Log-Weibull}\em
The marginal distribution of $\XX = (X_1,\ldots,X_d)$ is such that each of $\{\Lambda_i: i  = 1.\ldots,d\}$ are eventually strictly increasing and $\Lambda_i \in \RV(\alpha_i)$ for some $\alpha_i > 0.$ The joint distribution, when written in terms of the probability density $f_{\YY}(\cdot)$ of $\YY = \mv{\Lambda}(\XX),$ admits the form,
      \begin{align}
        f_{\mv{Y}}(\mv{y}) = p(\mv{y})\exp(-\varphi(\mv{y})),
        \label{pdf-Y}
       \end{align}
       where the functions $\varphi(\cdot),p(\cdot)$ satisfy the
       following: There exists a limiting function
       $I:\mathbb{R}^d_+ \rightarrow \mathbb{R}_+$ such that,
       \begin{align}
         n^{-1}\varphi(n\mv{y}_n) \rightarrow I(\yy) \quad \text{ and }  \quad
         n^{-\varepsilon}\log p(n\mv{y}_n) \rightarrow 0,
          \label{limiting-I}
       \end{align}
       for any sequence $\{\yy_n\}_{n \geq 1}$ of $\Real^d_+$
       satisfying $\yy_n \rightarrow \yy \neq \mv{0},$ and
       $\varepsilon > 0.$\label{item:Joint}
\end{assumption}
A wide variety of parametric and nonparametric multivariate distributions, including normal, exponential family, elliptical, log-concave distributions and Archimedian copula models satisfy Assumption~\ref{assume:Log-Weibull}. Marginal distributions which satisfy $\Lambda_i \in \RV(\alpha_i)$ include all distributions that are either Weibull-type heavy-tailed or possess lighter tails (such as exponential, normal, etc.). See \citeNP[Appendix B]{deo2021achieving} for further details and sufficient conditions directly in terms of the distribution of $\XX.$ 


\noindent{\textbf{Choice of the IS density.}} A cornerstone of  \VaR/\CVaR  \ estimation is the accurate estimation of the loss tail distribution, $1-F_L(u),$ for large values of $u$. In elementary examples, this is typically achieved  by choosing an IS density with features suitably mirroring the conditional distribution of $\boldsymbol{X}$ over $L(\xx)> u$ (see \citeNP[Section 4.2]{bucklew2013introduction}).  A central component in this endeavour is to utilize large deviations to identify the most likely way in which the loss $L(\XX)$ becomes large. For the broad family of losses and distributions specified by Assumptions \ref{assume:V} and \ref{assume:Log-Weibull} above, \citeNP{deo2021achieving}  show that the sequence of random vectors $\{t^{-1}\YY:t\geq 1\}$ satisfy (i) the following large deviations principle (LDP),
\begin{equation}\label{eqn:LDP}
    \lim_{t\to\infty} \frac{1}{t}\log \Prob\left( t^{-1}\YY  
\in A\right) = -\inf_{\yy\in A} I(\yy), \text{ for any Borel set $A$},
\end{equation}
and (ii) consequently, satisfy the tail asymptotic, 
\begin{equation}\label{eqn:LDP-DM}
\lim_{u\to\infty} \frac{1}{\Lambda(u^{1/\rho})}\log \Prob(L(\XX) > u) = -I^*,
\end{equation}
for some positive constant $I^\ast$ (see \citeNP[Theorems 3.3 and 4.1]{deo2021achieving}).

The lack of explicit dependence on parameter $u$ in the right-hand side of \eqref{eqn:LDP-DM} suggests that the concentration of the target conditional distribution of $(\XX \ \vert \ L(\XX) > v_\beta)$ may be approximated from the conditional samples of $(\XX\  \vert \ L(\XX) > l_\beta), $ where $l_\beta \ll v_\beta$ and $l_\beta \to\infty$ as the tail probability level $\beta \to 0.$  The requirements on $l_\beta$ ensure that the event $\{L(\XX) >l_\beta\},$ though rare by itself, is significantly less rare than the target event $\{L(\XX) > v_\beta\}$ and is observed more often in the samples. Letting $\l_\beta = v_\beta / r_\beta,$ the map $\mv{T}$ in Algorithm \ref{algo:CVaR-I.S.} suitably replicates these more frequent samples from the less rare region $\{L(\xx) > l_\beta\}$ onto the target set $\{L(\xx) > v_\beta\}.$  Specifically, the distribution of $\ZZ = \mv{T}(\XX)$  can be roughly written as approximating the conditional distribution of $\XX$ given $L(\XX)> v_\beta$ as in, 
\begin{equation}\label{eqn:heuristic}
    \frac{\log f_{\XX} (\xx) }{\log \Prob(L(\XX) > v_\beta)} \approx \frac{\log f_{\ZZ} (\xx)}{\log\Prob(L(\ZZ) > v_\beta)}, \text{ over $ \xx \in \{L(\xx) > v_\beta\}.$}
\end{equation}
\begin{example}
\textnormal{
To see \eqref{eqn:heuristic} by means of an example,
 suppose $\XX$ has a multivariate exponential distribution with density  $f_{\XX} (\xx) = g(\xx) \exp\left(-\|\xx\|_m\right)$, $\xx\in \R_+^d$, for some $g:\R^d_+\to\R_+$ and $m\in [1,\infty)$ (see \citeNP[Section 4]{lu1990some}). 
Changing variable to $\pp= v_\beta^{-1/\rho} \xx$ and letting $I^* = \inf_{L^*(\pp) \geq 1}\|\pp\|_m,$  we obtain
\begin{align*}
    \log f_{\XX} (v_\beta^{1/\rho}\pp)  = -v_\beta^{1/\rho}\|\pp\|_m (1+o(1)) \quad &\textrm{ and }  \quad \log f_{\ZZ}(v_\beta^{1/\rho} \pp) = -l^{1/\rho}_\beta\|\pp\|_m (1+o(1)),\\
    \log \Prob(L(\XX) > v_\beta) =  -v_\beta^{1/\rho} I^*  (1+o(1)) \quad &\textrm{ and }\quad  \log \Prob(L(\ZZ) > v_\beta) = -l^{1/\rho}_\beta I^* (1+o(1)),
\end{align*}
as $\beta \rightarrow 0,$ and over $\xx = v_\beta^{1/\rho}\pp$ in the region $\{\xx:L(\xx) > v_\beta\}.$}
\label{eg:multi-exp}
\end{example}
Indeed the approximating feature of $\mv{T}$ demonstrated in Example \ref{eg:multi-exp} can be shown to hold more generally for any $\XX$ satisfying Assumption~\ref{assume:Log-Weibull}; see \citeNP[Proposition 5.1]{deo2021achieving} for a precise statement of this self-structuring feature of the map $\mv{T}$ and the accompanying figures. 
The following asymptotic variance reduction guarantees for the proposed \VaR /\CVaR \ estimation are obtained as a consequence.  
\begin{theorem}\label{thm:CVaR-IS-LogWB}
Under  Assumptions~\ref{assume:V} and \ref{assume:Log-Weibull}, the IS estimators for \VaR \ and \CVaR \ returned by Algorithm \ref{algo:CVaR-I.S.} are asymptotically normal and offer the following variance reduction:
\[
\sqrt{n}(v_{\beta} - \hat{v}_{n,\beta}^{\mathrm{IS}}) \xrightarrow{\mathcal{D}} \sigma_{is,v}(\beta)N(0,1) \quad \text{ and } \quad \sqrt{n}(C_{\beta} - \hat{C}_{n,\beta}^{\mathrm{IS}}) \xrightarrow{\mathcal{D}} \sigma_{is,c}(\beta)N(0,1),
\] 
where the limiting variances, $\sigma^2_{is,v}(\beta)$  and  $\sigma^2_{is,c}(\beta),$ satisfy,
\[
\frac{ \sigma_{is, v}^2(\beta)}{ \sigma_{v}^2(\beta)}  = o(\beta^{1-\varepsilon}) \quad \text{ and } \quad  \frac{ \sigma_{is,c}^2(\beta)}{\sigma^2_c(\beta)}  = o(\beta^{1-\varepsilon}),
\]
as $\beta \rightarrow 0,$ when compared to the naive estimation variances $\sigma_{v}^2(\beta)$ and $\sigma^2_c(\beta)$  in \eqref{eqn:SAA}. 
\end{theorem}
Considering the proposed change of measure for the example of CVaR estimation, Theorem ~\ref{thm:CVaR-IS-LogWB} guarantees a sample complexity of $o(\beta^{-\varepsilon})$ as $\beta \searrow 0,$ where $\varepsilon > 0$ can be made arbitrarily small. Thus the asymptotic variance reduction is optimal when viewed in the logarithmic scale (see \shortciteNP{BJZWSC}). In contrast, naive estimation without any change of measure requires $\tilde{O}(\beta^{-1})$ samples. With the variance reduction guarantee holding for any choice of hyperparameter $h > 0,$ an effective $h$ can be chosen via cross-validation without incurring a change of scaling in sample complexity. The numerical experiments below demonstrate this by illustrating the relative insensitivity of variance reduction to various choices of $h.$


\section{NUMERICAL EXAMPLES}\label{sec:num} 
For a given loss $L(\cdot)$ and the random vector $\XX,$ we adopt the following procedure across all the experiments. 
Following Algorithm \ref{algo:CVaR-I.S.}, we take $n$ independent samples $\mv{X}_1,\ldots,\mv{X}_{n}$ to arrive at the IS c.d.f. estimate $\hat{F}_{n,L}^{\mathrm{IS}}(\cdot)$ in \eqref{eqn:FIS} and subsequently use it to arrive at the IS \VaR \ estimate $\vis := \inf\{ u : \hat{F}_{n,L}^{\mathrm{IS}} (u) \geq 1-\beta \}$ and the IS \CVaR \  estimate in \eqref{eqn:CVaR-comp-Is}. For  every choice of $\beta$ considered, the hyper-parameter $h$ is chosen by  performing cross-validation over the observed coefficient of variation. Each experiment involves computation of \CVaR\  as above from $n$ independent samples of $\XX$ and we report the relative root-mean square error = (root mean-square error of \CVaR\ observed across 50 independent experiments)/(average of \CVaR\ observed across 50 experiments). To enable comparison with naive estimation without IS, we also report its sample complexity for attaining the same precision offered by the IS algorithm.  We observe the following across the experiments: 1) the proposed IS has a significantly smaller relative error and a lower sample complexity when compared to estimation without any change of measure, and  2) the errors obtained using IS do not increase as the problem is made increasingly difficult by considering smaller values of $\beta.$ These observations align with the conclusions of Theorem \ref{thm:CVaR-IS-LogWB}. The specific details of the experiments are given below. 

\subsection{PERT Network:} 
\begin{figure}[h!]
      \begin{center}
      \begin{subfigure}{0.48\textwidth}
          \includegraphics[width=0.85\textwidth]{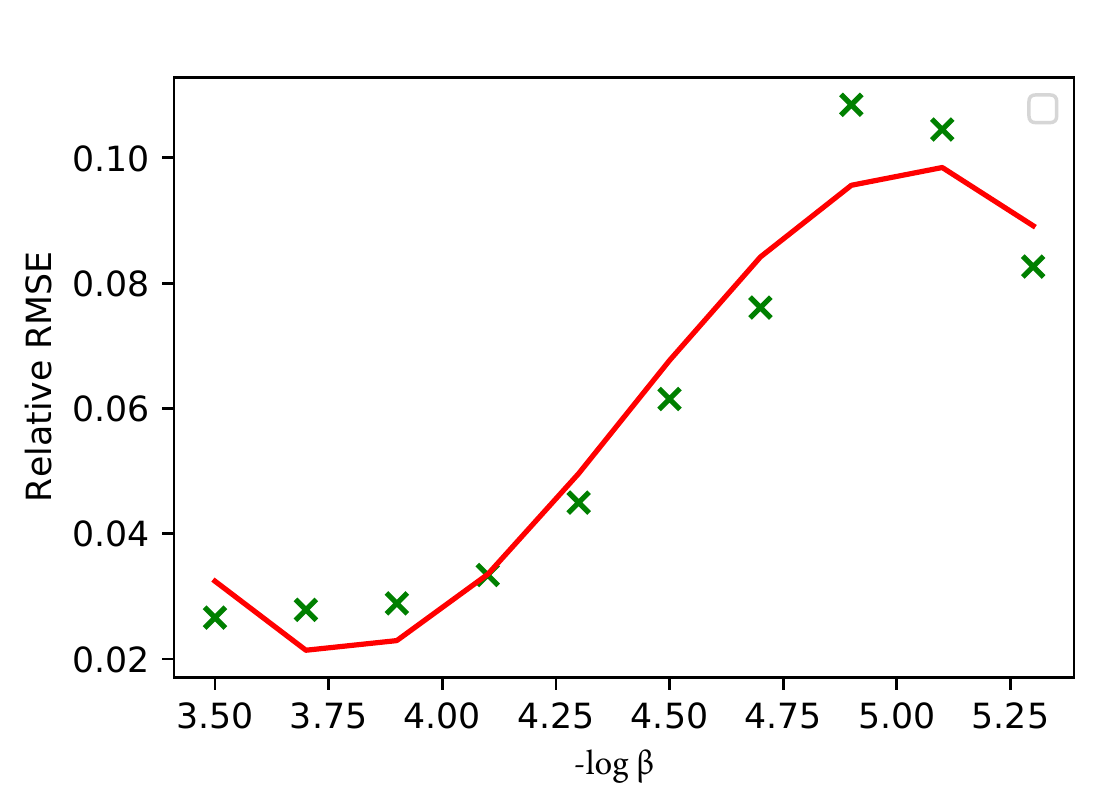}
          \caption{\small{Relative error in \CVaR \  estimation without IS}}
       \end{subfigure}
       \begin{subfigure}{0.48\textwidth}
          \includegraphics[width=0.85\textwidth]{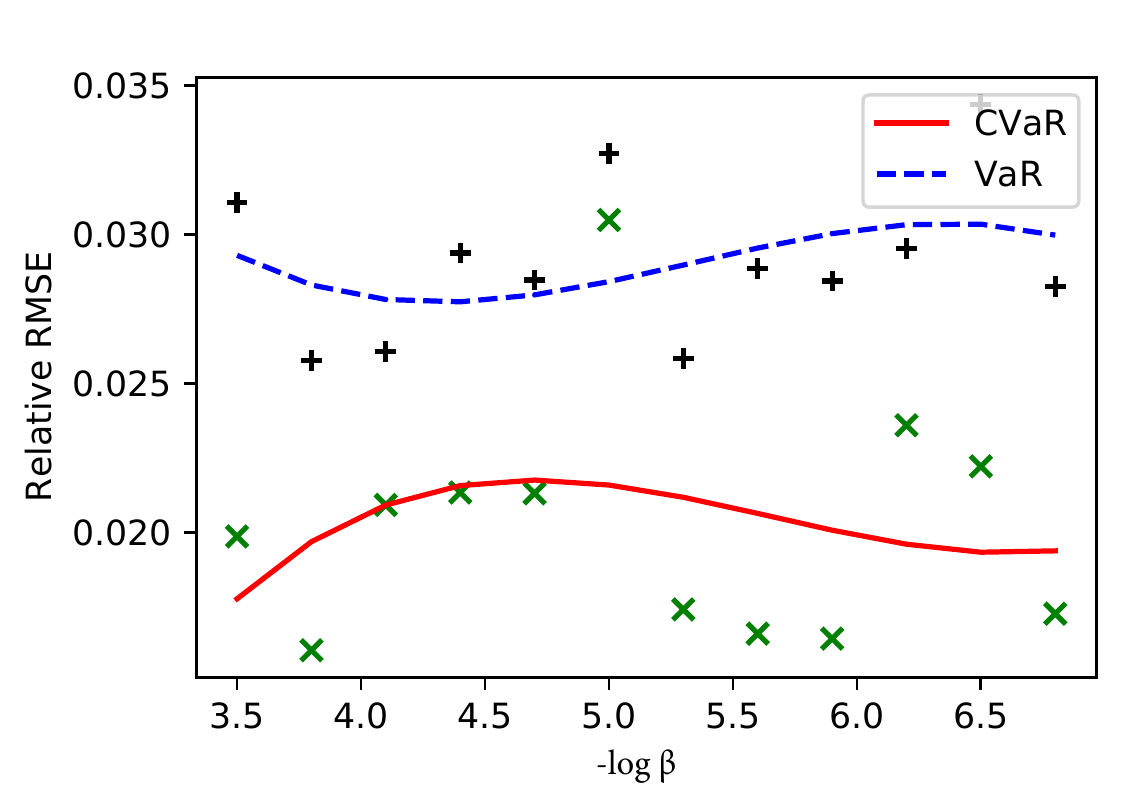}
          \caption{\small{Relative error in \VaR /\CVaR \  estimation using IS}}
       \end{subfigure}
      \end{center}
      \caption{{ Figure~\ref{fig:RMSE-PERT}(a) displays the relative RMSE in \CVaR \ estimation without IS.  The solid red curve is fit to the estimated relative RMSE from the sample estimates indicated by green crosses. Figure~\ref{fig:RMSE-PERT}(b) shows the relative errors in \VaR \  (blue fit line to black marks) and \CVaR\ (red fit line to green crosses) estimation using the IS scheme. The RMSE does not grow even as the tail probability level $\beta$ is made small.}\label{fig:RMSE-PERT} }
 \end{figure}

We consider a PERT network where the project completion time $L(\cdot)$ is generally written as the value of a mixed integer linear program. We consider an example with $d=7$ tasks and take $L(\xx) = x_1+x_7+\max\{x_5+\max\{x_2,x_3\}, x_6+\max\{x_4,x_3\}\}$. Here $L(\xx)$ is taken to be completion time of the PERT network when the individual task completion times realise the values $\xx$. To demonstrate performance for heavier than exponential delays, we assume that the marginal distribution of each delay is $F(x) = 1-\e^{-x^{0.5}}$ and their joint dependence is through a Gaussian copula whose correlation matrix is given by 
\begin{equation}\label{eqn:corr-GC}
R_{i,j} =\begin{cases}
0.1 &\text{ if } |i- j|=1,\\
1 & \text{ if } i=j,\\
0 & \text{ other-wise.}
\end{cases}.
\end{equation}
In each experiment, we take $n=10^3$ samples to compute \VaR /\CVaR \ using the IS estimator. We plot the observed root mean square errors  (observed across 50 independent experiments) in Figure \ref{fig:RMSE-PERT}  as a function of the tail probability level $\beta\in(10^{-7},10^{-3})$. The  parameter $h$ is selected as $h(\beta)  = 2 -0.6\log \beta$. Figure~\ref{fig:RMSE-PERT}(a)  details the results. Contrast this to estimation without IS which requires $\approx 2\times 10^{5}$ samples to attain a relative error similar to the IS scheme at $\beta = 10^{-3.5}$ (see Figure~\ref{fig:RMSE-PERT}(b)).
 \subsection{Linear portfolios:} 
\begin{figure}[t]
      \begin{center}
      \begin{subfigure}{0.48\textwidth}
          \includegraphics[width=0.85\textwidth]{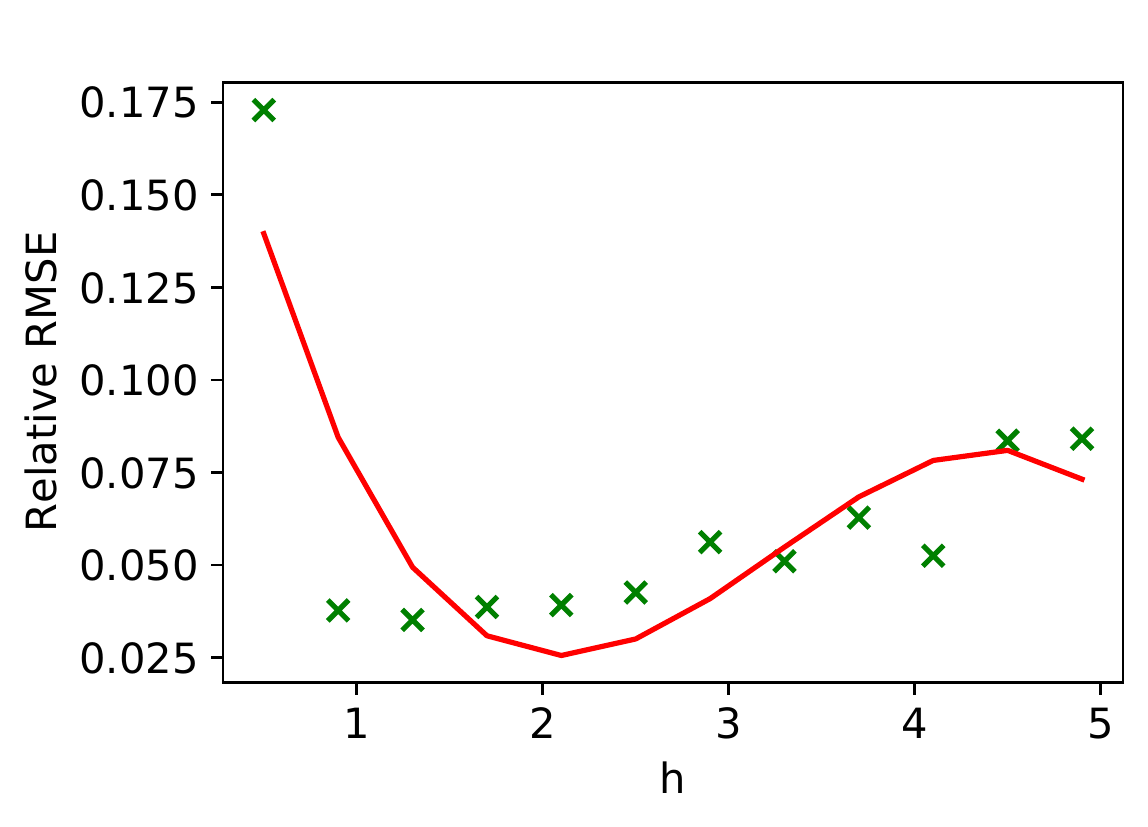}
          \caption{\small{Cross validation over $h$}}
       \end{subfigure}
       \begin{subfigure}{0.48\textwidth}
          \includegraphics[width=0.85\textwidth]{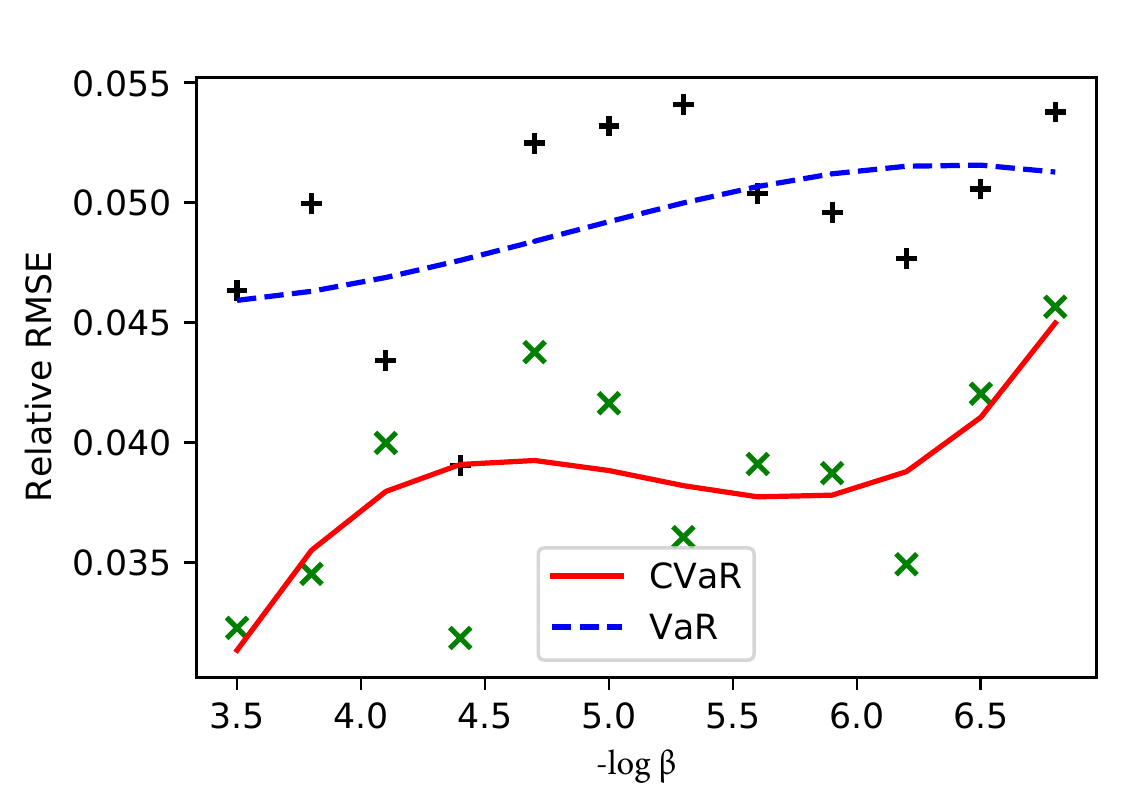}
          \caption{\small{Relative error in \VaR / \CVaR \  estimation using IS}}
       \end{subfigure}
      \end{center}
      \caption{{Figure~\ref{fig:RMSE-Portfolio}(a) displays the relative RMSE in \CVaR \  computation against the parameter $h$ (for the target level $\beta=10^{-6}$) and Figure~\ref{fig:RMSE-Portfolio}(b) shows the relative errors observed for $\beta \in [10^{-3.5}, 10^{-7}].$  The solid curves are fit to the observed relative errors  in \VaR\  (black marks) and \CVaR\ (green crosses) estimation. In Figure~\ref{fig:RMSE-Portfolio}(b), $h=2.6$. 
      }\label{fig:RMSE-Portfolio}}
 \end{figure}

We consider the equally weighted linear portfolio loss $L(\xx) = \mv{1}^\intercal{\xx}$ in this example. To illustrate performance in a case where the marginal distributions of the components of $\XX \in \mathbb{R}^{10}$ are different, we consider the marginal c.d.f.s $F_i(x) = P(X_i \leq x) = 1-\e^{-x^{\alpha_i}}$ where $\alpha_i =0.9$ for $1\leq i\leq 5$, and $\alpha_i=1.1$ for $6\leq i\leq 10$.  Dependence among the components of $\XX$ is introduced through a Gaussian copula for which the correlation matrix $R$ is specified by the off-diagonal entries $[R]_{i,j} = 0.1$ for
$i\neq  j$, and diagonal entries $[R]_{i,i} =1$ for $i \in \{1,\ldots,10\}$.   Figure \ref{fig:RMSE-Portfolio}(a) below presents
details on cross-validation by plotting the relative error of
estimation observed for different
choices of hyper-parameter $h$ considered. With the relative RMSE staying less than $5\%$ throughout the interval $h \in (1.5 ,3.5),$ we note that the error reduction is robust and the estimator variance is relatively less sensitive to the choice of parameter $h.$ Notice from  Figure~\ref{fig:RMSE-Portfolio}(b) that a relative error between $3\%-4\%$ is obtained with only $n=10^3$ samples upon use of the IS algorithm, and that this error is constant even as the target level $\beta$ is varied from $10^{-3.5} $ to $10^{-7}$. Note that to obtain a $3\%$ relative error at level $\beta=  10^{-3.5}$, estimation without IS requires $n\approx 2\times 10^5$ samples. 
\subsection{Forest fires data-set}  
\begin{figure}[t]
      \begin{center}
      \begin{subfigure}{0.48\textwidth}
          \includegraphics[width=0.85\textwidth]{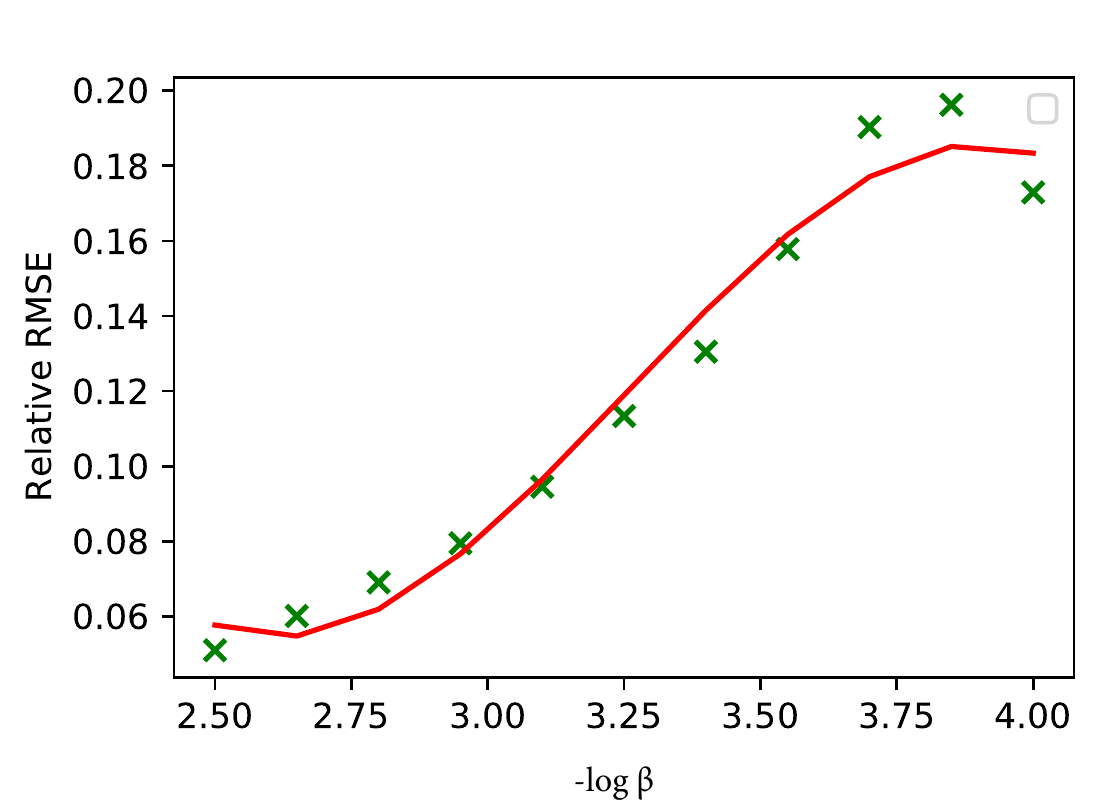}
          \caption{\small{Relative error in \CVaR \ estimation without IS}}
       \end{subfigure}
       \begin{subfigure}{0.48\textwidth}
          \includegraphics[width=0.85\textwidth]{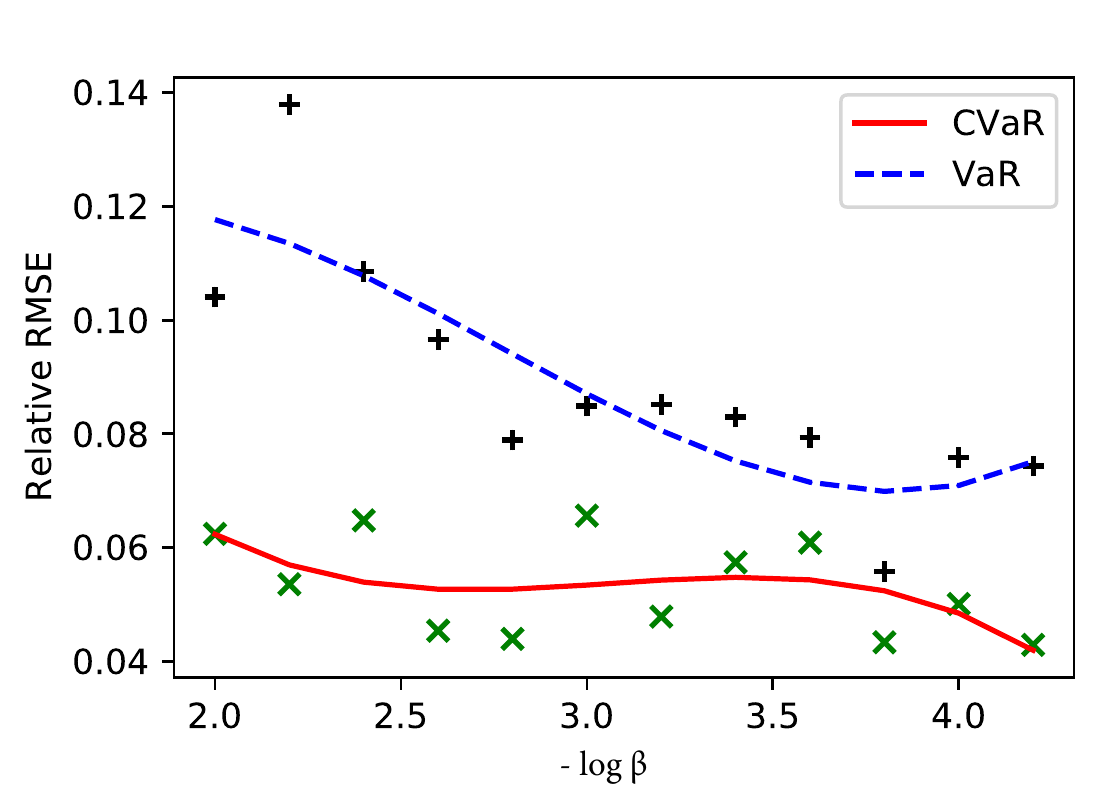}
          \caption{\small{Relative error in \VaR /\CVaR \  estimation with IS}}
       \end{subfigure}
      \end{center}
      \caption{{Figure~\ref{fig:RMSE-NN}(a) displays the relative RMSE in \CVaR \  computation without IS using $10^4$ samples. Figure~\ref{fig:RMSE-NN}(b) displays the relative errors in IS based \VaR /\CVaR \ computation with the parameter $h=4.6$. In each of the figures, black and green marks respectively denote estimated \VaR \ and  \CVaR \ values respectively.
      } \label{fig:RMSE-NN} }
 \end{figure}

We consider a loss trained from the  \hyperlink{https://archive.ics.uci.edu/ml/datasets/forest+fires} {forest fires dataset} used in \citeNP{cortez2007data} in this example.  The input covariates $\XX$ consist of climatic factors such as wind speed, daily rainfall, temperature, humidity etc. The output $L(\XX)$ is the area of forest fires, in hectares, corresponding to the respective climatic data. We train a deep neural network (DNN) network to learn the function $L_{\mv{\theta}}(\cdot)$, which maps the covariates to the log of the area of the forest fire. The DNN has one hidden layer consisting of 12 neurons with ReLU link. The parameters $\mv{\theta}$ are learnt via stochastic gradient descent.  We consider the following example distribution for covariates $\XX$ for the sake of the experiment: The marginal distribution of the components are given by $F(x) = 1- \e^{-x^{0.6}}$ and the dependence structure is informed via a Gaussian copula whose correlation matrix is given as in \eqref{eqn:corr-GC}. For the purpose of this experiment, we choose $n=517$ in \eqref{eqn:CVaR-comp-Is} to match the size of the input data-set. As before, we cross validate over the parameter $h$ (see Figure~\ref{fig:RMSE-NN}(a)), and then using $h=4.6$ in Algorithm~\ref{algo:CVaR-I.S.}, jointly estimate \VaR\ / \CVaR \ for $\beta\in(10^{-4.5},10^{-2})$. Figure~\ref{fig:RMSE-NN}(b) gives the result of our experiment. It is worthwhile to note that although the loss $L_{\mv{\theta}}(\cdot)$ is a black box, our algorithm still produces estimates of \VaR /\CVaR \ with a small relative error (4-6\% for \CVaR \ and 7-10\% for \VaR ). Contrast this to MC estimation, which requires $ n \approx 7\times 10^{3}$ samples to give a relative error of $4\%$ in \CVaR \ estimation at $\beta=10^{-2}$.
  \section{PROOF OF THEOREM \ref{thm:CVaR-IS-LogWB}} \label{sec:proofs}
For ease of presentation, we focus on variance reduction in CVaR estimation and assume that $\XX $ has identical marginals (that is, $\Lambda_i=\Lambda$ for all $i$). The proof for the case where $\XX$ has heterogeneous marginals can be similarly accomplished by introducing a vector for capturing differing relative tail heaviness as in the results in \citeNP{deo2021achieving}.
To begin, we recall \citeNP[Corollary 2]{sun2010asymptotic}, as applicable to our IS estimator: 
\begin{equation}\label{eqn:CVaR-CLT}
\sqrt{n}(\cis - C_\beta) \xrightarrow{\mathcal{D}} \sigma_{is,c}(\beta)N(0,1)
\end{equation}
where $\beta^{2}\sigma_{is,c}^2(\beta) = {{\textrm{Var}}\left[(L(\ZZ)-v_\beta)^+\mathcal{L}_R\right]}$ and the likelihood ratio $\mathcal{L}_R$ is defined as in \eqref{eq:LR}. To present the main ideas in deconstructing the above variance term, we postpone the verification of the technical conditions required for applying \citeNP[Corollary 2]{sun2010asymptotic} towards the end of this section. 

For any $\mv{a}\in \R^d_+$ and $r>0$, let $B_r(\mv{a}) = \{\yy\in \R^d_+: \|\mv{a}-\mv{\yy}\|_\infty\leq r\}$ be a ball of radius $r,$ centred at $\mv{a},$ under the $\|\cdot\|_\infty$ norm.   Denote $B_{r}(\mv{0})$ by $B_r$.  Define $\mv{q} = \mv{\Lambda}^{-1}$ be the component-wise inverse, $\mv{\psi}_\beta = \mv{\Lambda} \circ \mv{T}^{-1} \circ \qq$  and $t(\beta)=\Lambda(v^{1/\rho}_\beta)$. Let $\lambda_i(x) = f_{X_i}(x)/(1-F_{X_i}(x))$ denote the hazard rate of $X_i$ and $E[X;A] = E[XI(A)].$ Define $L_\beta(\pp) = [v_\beta]^{-1} L(\qq(t(\beta)\pp))$. Finally, let $\YY_\beta = [t(\beta)]^{-1}\YY$.  For notational convenience, let $M_{2,\beta}$ denote the second moment of $[\mathcal{L}_R(L(\ZZ) - v_\beta)^+]$.  For $A\subseteq \R_+^d$, let $\chi_{A}(\cdot)$ denote its characteristic function; that is, $\chi_A(\xx) =\infty$ if $\xx\not\in A$ and $\chi_A(\xx) = 0$ if $\xx\in A$. Further, for a function $f:\R^d_+\to \R$ and $a \in \Real,$ let $\Lev_a^+(f) = \{\xx\in \R^d_+ :f(\xx)\geq a\}$ denote the super-level set of $f.$ Let $\fLD(\xx) = L^*(\xx^{1/\alpha})$. 

With this notation, see that $\YY_\beta= [t(\beta)]^{-1} \YY= [t(\beta)]^{-1}\mv{\Lambda}(\XX)$. Changing variables from $\XX $ to $\YY_\beta $ in the expectation below  (see (EC.16) onward in the proof of Lemma EC.6 of \citeNP{deo2021achieving} for detailed steps in a similar change of variables exercise), we obtain 
\begin{align}
M_{2,\beta}= \Expc \left[(L(\XX) - v_{\beta})^2 \frac{f_{\XX}(\XX)}{{f}_{\XX}(\mv{T}^{-1}(\XX))} J(\mv{T}^{-1}(\XX)); L(\XX) \geq v_\beta \right] = \Expc\left[\exp(-t(\beta) F_\beta(\YY_\beta))\right]\label{eqn:M2-firststep}
\end{align}
\begin{subequations}
\begin{equation}
    \text{where } F_\beta(\pp) = a_{\beta}(\pp) + b_{\beta}(\pp) +c_\beta(\pp) - 2d[t(\beta)]^{-1}\log t(\beta) + \chi_{\Lev_1^+(L_\beta)}(\pp)]   
\end{equation}
\begin{equation}\label{eqn:abeta}
\text{ where  } 
    a_\beta(\pp) = [t(\beta)]^{-1} \left[ \log f_{\YY}(\mv{\psi}_u(t(\beta)\pp)    - \log f_{\YY}(t(\beta)\pp) \right], \text{ and } 
\end{equation}
\begin{equation}\label{eqn:bbeta}
 b_\beta(\pp) = [t(\beta)]^{-1} \left[\sum_{i=1}^d \left[
    \log\lambda_i(\mv{T}^{-1}_i(\mv{q}(t(\beta)\pp))) -
    \log\lambda_i(q_i(t(\beta)p_i))\right] -
    \log J(\mv{T}^{-1}(\mv{q}(t(\beta)\pp)))\right] \text{ and }
\end{equation} 
\begin{equation}\label{eqn:cbeta}
 c_\beta(\pp) = -2[t(\beta)]^{-1} \log(L(\qq(t(\beta) \pp) -v_{\beta})  . 
\end{equation} 
\end{subequations}
Notice that $c_\beta(\pp)$ is well defined for all $\pp\in \Lev_1^+(L_\beta)$. Observe that from \eqref{eqn:LDP-DM}, $v_\beta \sim q^{\rho}(-I^{*}\log \beta)$. Next, recall that $\Lambda\in \RV(\alpha)$ and that $\Lambda^{-1} =q$. Hence, from \citeNP[Proposition B.1.9 (viii)]{de2007extreme}, $q \in \RV(1/\alpha)$. Therefore, $r_\beta/v_{\beta} \to 0$ and $r_\beta\to\infty$ as $\beta\to\ 0$.  Hence, the following conclusions of \citeNP[Lemmas A.8-A.11 and Corollary A.3]{deo2021achieving} hold:  for $\varepsilon, r>0$, for all sufficiently small enough $\beta$,

\begin{subequations}
\begin{equation}\label{eqn:b-beta-bound}
\sup_{\pp\in \R^d_+} b_{\beta}(\pp) \geq 0
\end{equation}
\begin{equation}\label{eqn:0notincl}
\Lev_1^+(L_\beta)\cap B_{\delta_1} = \emptyset \text{ for some $\delta_1>0$,} 
\end{equation}
\begin{equation}\label{eqn:a-beta-bound}
 a_\beta(\pp) \geq I(\pp) +o(1) \text{ uniformly over $\pp \in  \Lev_1^+(L_\beta) \cap B_r$, and } \liminf_{\beta\to 0 } \chi_{\Lev_1^+(L_{\beta})}(\pp_{\beta}) \geq \chi_{\Lev_1^+(\fLD)}(\pp) 
\end{equation}
\end{subequations}
whenever $\pp_\beta\to\pp.$ Let $\hat{\pp} = \pp/\Vert \pp\Vert_\infty$ be the unit vector in the direction of $\pp$.  Rewrite 
\begin{align*}
L(\qq(t(\beta)\pp ) ) &=\frac{ L\left(\frac{\qq(t(\beta)\|\pp\|_\infty \hat{\pp})}{q(t(\beta)\|\pp\|_\infty)} q(t(\beta)\|\pp\|_\infty)\right)}{q^\rho(t(\beta)\|\pp\|_\infty)} q^\rho(t(\beta)\|\pp\|_\infty) = L^*(\hat{\pp}^{1/\alpha})q^\rho(t(\beta)\|\pp\|_\infty) (1+o(1)),
\end{align*}
uniformly over  $\|\pp\|_\infty  \geq \delta$; the second equality in the above is obtained upon noting that $q\in RV(1/\alpha)$, and using the continuous convergence of $L(\cdot)$ as specified in Assumption~\ref{assume:V}. Further, as $x\to\infty$, for any $\varepsilon>0$,
\[
\frac{\log q(x)}{\varepsilon x} \to 0 \text{ \quad see \citeNP[Proposition B.1.9 (1)]{de2007extreme}}.
\]
  Therefore, \eqref{eqn:0notincl} suggests that  uniformly over $\Lev_1^+(L_\beta)$, ${\log L(\qq(t(\beta))\pp)} \leq {\varepsilon t(\beta)\|\pp\|_\infty},$ for all $\beta$ sufficiently small. Further, since $t(\beta) \leq \exp\left(\varepsilon t(\beta)\right)$ for all small enough $\beta,$
\begin{equation}\label{eqn:smaller-term}
 c_{\beta}(\pp) \geq -\varepsilon\|\pp\|_\infty \text{ uniformly over $\Lev_1^+(L_\beta)$.}
\end{equation}
Now for any $\varepsilon>0,$ from the bounds in \eqref{eqn:b-beta-bound}, \eqref{eqn:a-beta-bound} and \eqref{eqn:smaller-term}, one obtains that whenever $\pp_\beta\to\pp$ as $\beta\to 0 $,
\[
\liminf_{\beta\to 0 } F_{\beta}(\pp_\beta) \geq I(\pp) -\varepsilon\|\pp\|_\infty +\chi_{\Lev_1^+(\fLD)}(\pp).
\]
Noting that $\YY_\beta$ satisfies an LDP with rate function $I(\cdot)$, an application of the general Varadhan's integral lemma (see \citeNP[Theorem 2.2]{LD_Varadhan}) yields, 
\begin{equation}\label{eqn:crit-inf}
\limsup_{\beta\to 0}[t(\beta)]^{-1} \log M_{2,\beta}   \leq  -\inf_{\pp\in \Lev_1^+(\fLD)} \left[2I(\pp) -\varepsilon\|\pp\|_\infty\right].    
\end{equation}
Since $\YY$ has standard exponential marginals, $I(\pp) \geq \|\pp\|_\infty$  for all $\pp$  (see \citeNP[Lemma 3.4 (d)]{deo2021achieving}). The infimum in \eqref{eqn:crit-inf} therefore occurs in a compact set. As $\varepsilon>0$ above is arbitrary, we have
\begin{equation}
    \limsup_{\beta\to 0}[t(\beta)]^{-1} \log M_{2,\beta}  \leq -2\inf_{\pp\in\Lev_1^+(\fLD)}I(\pp) = -2I^*.
\end{equation}
Next, as a consequence of \eqref{eqn:LDP-DM}, $(1+o(1))I^*t(\beta) =  -\log P(L(\XX) > v_\beta)= -\log\beta$ as $\beta\to 0$. With $t(\beta) = -  (1/I^\ast + o(1))\log \beta,$ we have $\log M_{2,\beta} \leq (2-\delta) \log \beta.$ With the choice of $\delta > 0$ being arbitrary, we therefore have $M_{2,\beta} = o(\beta^{2-\delta})$ for any $\delta>0$. Finally, for the Monte Carlo estimator without change-of-measure, $\beta^2\sigma^2_c(\beta) = \Expc\left([(L(\XX) - v_{\beta})^+]^2    \right) - \left(\Expc\left( (L(\XX) - v_{\beta})^+\right)\right)^2$.
Notice that 
\begin{equation}\label{eqn:HighVar}
    \Expc\left([(L(\XX) - v_{\beta})^+]^2 \right)    = \int_{L(\xx) \geq v_\beta} (L(\xx) - v_{\beta})^2 f_{\XX}(\xx) d\xx \ \geq  P( L(\XX)  \geq v_{\beta}+1) \ = \beta (1 + o(1)).
\end{equation}
Further, notice that following the analysis from \eqref{eqn:M2-firststep}, for any $\delta>0$, $\Expc\left( (L(\XX) - v_{\beta})^+\right) \leq \beta^{1-\delta}$. Hence, $\left(\Expc\left( (L(\XX) - v_{\beta})^+\right)\right)^2 = o(\Expc\left([(L(\XX) - v_{\beta})^+]^2 \right) ) $. Thus, we have that for all $\delta>0$,
\[
\frac{\sigma_{is,c}^2(\beta)}{\sigma^{2}_c(\beta)} = o(\beta^{1-\delta}).\qed
\]
\noindent \textbf{Verification of the conditions of \citeNP[Corollaries 1 and 2 ]{sun2010asymptotic}:} Here we perform the pending verification of   \citeNP[Assumption 2]{sun2010asymptotic}. Notice that  existence of  $f_L(\cdot)$  automatically implies that \citeNP[Assumption 1]{sun2010asymptotic} holds, which is a sufficient condition for the central limit theorem to hold.  Fix any $p>2$. Notice that using a similar change of variables arguments as in the beginning of the proof (with $\tilde{\Expc}$ denoting expectation under the IS measure),  $\widetilde{\Expc}\left( \mathcal{L}_R^{p}\mathbf{I}(L(\XX) \geq v_\beta+\varepsilon) \right)$ is bounded above by
\begin{align}\label{eqn:verifyHS}
   \Expc\left[\exp(-(p-1)t(\beta) F_\beta(\YY_\beta) )  \right],
\end{align} 
for any $\varepsilon>0$. Following \eqref{eqn:b-beta-bound} through to \eqref{eqn:a-beta-bound}, 
$
\tilde{\Expc}\left(\mathcal{L}_R^{p} \mathbf{I}(L(\XX) \geq v_\beta +\varepsilon)\right)  \leq  \exp(t(\beta)p\varepsilon) \text{ for any $p>2$}$. \qed

 
 \section*{AUTHOR BIOGRAPHIES}
{\footnotesize  {\bf ANAND DEO} is a Senior Research Assistant at Singapore University of Technology and Design. His research interests span applied probability, quantitative risk management, operations research, and machine learning. Formerly, he was a PhD student at the Tata Institute of Fundamental Research, Mumbai. His e-mail address is \email{deo\_avinash@sutd.edu.sg}.}

\noindent {\footnotesize  {\bf KARTHYEK MURTHY} is an Assistant Professor in Singapore University of Technology and Design. His research centers around building models and methods for incorporating competing considerations such as risk, robustness, and fairness in data-driven optimization problems affected by uncertainty.  Before joining SUTD, he was a postdoctoral researcher at IEOR deparment, Columbia University and a PhD student at the Tata Institute of Fundamental Research, Mumbai. His e-mail address is \email{karthyek\_murthy@sutd.edu.sg}.}

 \section*{ACKNOWLEDGEMENTS} Support from Singapore Ministry of Education grant MOE2019-T2-2-163 is gratefully acknowledged. 

 \bibliographystyle{wsc}
\bibliography{demobib}
\end{document}

%% file: wscbib.tex
\makeatletter
\let\@internalcite\cite
\def\cite{\def\@citeseppen{-1000}%
    \def\@cite##1##2{(##1\if@tempswa , ##2\fi)}%
    \def\citeauthoryear##1##2##3{##1 ##3}\@internalcite}
\def\citeNP{\def\@citeseppen{-1000}%
    \def\@cite##1##2{##1\if@tempswa , ##2\fi}%
    \def\citeauthoryear##1##2##3{##1 ##3}\@internalcite}
\def\citeN{\def\@citeseppen{-1000}%
    \def\@cite##1##2{##1\if@tempswa, ##2)\else{}\fi}%
    \def\citeauthoryear##1##2##3{##1 (##3)}\@citedata}
\def\citeA{\def\@citeseppen{-1000}%
    \def\@cite##1##2{(##1\if@tempswa , ##2\fi)}%
    \def\citeauthoryear##1##2##3{##1}\@internalcite}
\def\citeANP{\def\@citeseppen{-1000}%
    \def\@cite##1##2{##1\if@tempswa , ##2\fi}%
    \def\citeauthoryear##1##2##3{##1}\@internalcite}
\def\shortcite{\def\@citeseppen{-1000}%
    \def\@cite##1##2{(##1\if@tempswa , ##2\fi)}%
    \def\citeauthoryear##1##2##3{##2 ##3}\@internalcite}
\def\shortciteNP{\def\@citeseppen{-1000}%
    \def\@cite##1##2{##1\if@tempswa , ##2\fi}%
    \def\citeauthoryear##1##2##3{##2 ##3}\@internalcite}
\def\shortciteN{\def\@citeseppen{-1000}%
    \def\@cite##1##2{##1\if@tempswa, ##2\else{}\fi}%
    \def\citeauthoryear##1##2##3{##2 (##3)}\@citedata}
\def\shortciteA{\def\@citeseppen{-1000}%
    \def\@cite##1##2{(##1\if@tempswa , ##2\fi)}%
    \def\citeauthoryear##1##2##3{##2}\@internalcite}
\def\shortciteANP{\def\@citeseppen{-1000}%
    \def\@cite##1##2{##1\if@tempswa , ##2\fi}%
    \def\citeauthoryear##1##2##3{##2}\@internalcite}
\def\citeyear{\def\@citeseppen{-1000}%
    \def\@cite##1##2{(##1\if@tempswa , ##2\fi)}%
    \def\citeauthoryear##1##2##3{##3}\@citedata}
\def\citeyearNP{\def\@citeseppen{-1000}%
    \def\@cite##1##2{##1\if@tempswa , ##2\fi}%
    \def\citeauthoryear##1##2##3{##3}\@citedata}
%
%
%
\def\@citedata{%
    \@ifnextchar [{\@tempswatrue\@citedatax}%
                  {\@tempswafalse\@citedatax[]}%
}

\def\@citedatax[#1]#2{%
\if@filesw\immediate\write\@auxout{\string\citation{#2}}\fi%
  \def\@citea{}\@cite{\@for\@citeb:=#2\do%
    {\@citea\def\@citea{, }\@ifundefined
       {b@\@citeb}{{\bf ?}%
       \@warning{Citation `\@citeb' on page \thepage \space undefined}}%
{\csname b@\@citeb\endcsname}}}{#1}}%

%
\def\@citex[#1]#2{%
\if@filesw\immediate\write\@auxout{\string\citation{#2}}\fi%
  \def\@citea{}\@cite{\@for\@citeb:=#2\do%
    {\@citea\def\@citea{; }\@ifundefined
       {b@\@citeb}{{\bf ?}%
       \@warning{Citation `\@citeb' on page \thepage \space undefined}}%
{\csname b@\@citeb\endcsname}}}{#1}}%

%
\def\@biblabel#1{}
\makeatother



\newdimen\bibindent
\bibindent=0.0em
\def\thebibliography#1{\section*{\refname}\list
   {}{\settowidth\labelwidth{[#1]}
   \leftmargin\parindent
   \itemindent -\parindent
   \listparindent \itemindent
   \itemsep 0pt
   \parsep 0pt}
   \def\newblock{}
   \sloppy
   \sfcode`\.=1000\relax}